\documentclass[fleqn,10pt]{wlscirep}
\title{Refractive index measurements of multicellular tumour spheroids using optical coherence tomography: dependence on growth phase and size}

\usepackage{graphicx}
\usepackage{textcomp}
\usepackage{multirow}

\author[1,2]{Neelam Hari}
\author[3]{Priyanka Patel}
\author[4]{Jacqueline Ross}
\author[3]{Kevin Hicks}
\author[1,2,*]{Fr\'{e}d\'{e}rique Vanholsbeeck}
\affil[1]{Department of Physics, University of Auckland, Auckland 1010, New Zealand}
\affil[2]{The Dodd-Walls Centre for Photonic and Quantum Technologies, New Zealand}
\affil[3]{Auckland Cancer Society Research Centre, University of Auckland, Auckland, New Zealand}
\affil[4]{Biomedical Imaging Research Unit, Department of Anatomy and Medical Imaging, University of Auckland, Auckland, New Zealand}
\affil[*]{f.vanholsbeeck@auckland.ac.nz}

\keywords{refractive index, optical coherence tomography, multicellular spheroids, confocal microscopy, optical clearing agents}

\begin{abstract}
Knowledge of optical properties, such as the refractive index (RI), of biological tissues is important in optical imaging, as they influence the distribution and propagation of light in tissue. To accurately study the response of cancerous cells to drugs, optimised imaging protocols are required. This study uses a simple custom-built spectral domain optical coherence tomography (OCT) system to conduct RI measurements of multicellular spheroids, three-dimensional \textit{in-vitro} culture systems, of the cell line HCT116. The spheroid RIs are compared to study the effect of growth time. To improve confocal microscopy imaging protocols, two immersion media (glycerol and ScaleView-A2) matching the spheroid RIs were trialled, with the aim to reduce the RI mismatch between the spheroid and the immersion medium and thus improving imaging depth with confocal microscopy. ScaleView-A2 (n = 1.380) aided in achieving greater depths of imaging of the multicellular spheroids under confocal microscopy. This improvement in imaging depth confirmed the utility of our RI measurements, proving the promising outlook of OCT as a complementary tool to microscopy in cancer research. 
\end{abstract}

\begin{document}

\flushbottom
\maketitle
\thispagestyle{empty}

\section*{Introduction}

In cancer research, three-dimensional tumour cell cultures, also known as multicellular spheroids, serve as a biochemical and morphological \textit{in-vitro} research model for \textit{in-vivo} tumours \cite{MCTS1}. In comparison to two-dimensional \textit{in-vitro} cell models, multicellular spheroids are effective for studying cancer drug delivery with the advantage of resembling \textit{in-vivo} tumours in terms of cellular environment with respect to oxygen and pH gradients, as well as structural heterogeneity, thus contributing to the formation of quiescent, anoxic, hypoxic, and necrotic cell subpopulations \cite{casciari1992variations}. The \textit{in-vitro} multicellular spheroid exhibits a closely-packed, spherical geometry of cells with a concentric arrangement of actively proliferating cells at the periphery, an intermediate layer of quiescent cells, and an emerging central core of dead cells deep within the spheroid \cite{mueller2000tumor}. The rapid multiplication of tumour cells leads to a deprivation of oxygen within the inner cells.  Beyond a critical size ($\mathrm{>500 \mu m}$), most spheroids develop hypoxia leading to a necrotic core surrounded by a viable rim of cells \cite{MCTS1}. Where these conditions prevail, they contribute to a type of cell death, necrosis.\\

Hypoxic regions have an important role in tumour progression and are considered a target to exploit in cancer therapy \cite{targ1}. Recent studies have been conducted with spheroids to study the tumour cell response to hypoxia-activated prodrugs (HAP) \cite{Mao2}, which become activated at the oxygen-poor core of solid tumours. Spheroids are treated with hypoxia probes and live/dead stains to quantify the amount of hypoxia and assess the viability of cells using confocal microscopy. However, the constituents of biological tissue and lack of transparency scatter light, reducing the performance of confocal microscopy as depth into the specimen increases. In addition, refractive index (RI) mismatches limit the ability to image the entire depth of spheroids, resulting in low contrast, reduced spatial resolution with depth, and spherical aberrations \cite{Diaspro:02}. \\

A hypoxia marker has been developed at the Auckland Cancer Society Research centre (ACSRC), which is used with a click chemistry fluorescent probe. This allows the imaging of whole intact spheroids by confocal microscopy \cite{hong2018cellular}. Histology has demonstrated that penetration of the probe is not an issue as staining of whole intact spheroids prior to processing still demonstrated hypoxic cores. Widefield fluorescence microscopy is only suitable for imaging thin samples as thicker samples result in the problem of out-of-focus blur as fluorescence from multiple planes within the sample is captured in each image by the camera thereby providing a false impression of the location of staining. Therefore, in order to determine the distribution of a fluorescent dye throughout a multicellular spheroid with widefield fluorescence microscopy, embedding and serial sectioning is required. Confocal microscopy enables imaging of thicker samples with improved resolution and contrast through its ability to acquire optical sections as only in-focus light is captured by the detector. However, confocal microscopy still has limitations, as imaging deep into samples can be problematic due to light scatter from biological structures and RI mismatch. Depth of imaging can be improved by matching the RI of the sample with the immersion medium for the objective lens, thereby substantially reducing light scattering and refraction. The ability to measure the RI of the sample allows RI matching to be achieved for confocal microscopy resulting in improved image quality and depth. \\

Immersion media, such as glycerol and ScaleView-A2 are examples of optical clearing agents often used in optical microscopy. Optical clearing agents can improve imaging depth by reducing the biological tissue’s scattering and absorption, and thereby increasing light transmission \cite{welch2011optical}. This is especially important in optical imaging, as it enhances the ability to visualise structures at greater depths. The decreased scattering is achieved via mechanisms such as the replacement of tissue fluid with media of higher RI, dehydration, and collagen dissociation \cite{welch2011optical}.\\

Optical coherence tomography (OCT) is a technique which is capable of imaging the entirety of the spheroid, while also providing morphological and physiological information about it. It is label-free, using the intrinsic contrast provided by the sample, and also, non-invasive, which is advantageous in the realm of cancer research. Previous studies have shown that diameter and attenuation can be measured with OCT \cite{sharma2007imaging, huang2017optical}. Huang et al. (2017) conducted an impressive study with OCT detecting the necrosis within the multicellular spheroids using a label-free intrinsic optical attenuation contrast near the centre of the spheroid \cite{huang2017optical}. Furthermore, they took three-dimensional OCT images to visualise the morphology of the spheroids and study their growth kinetics and quantify their volumes using a voxel-based approach. While this study has provided valuable information of the potential of OCT as a high-throughput imaging system, the authors have not measured the multicellular spheroid RI. \\ 

The RI of tissue has been studied since the 1950s \cite{LiuRefractive} and is an important optical property of tissue in optical imaging techniques, such as confocal microscopy and OCT. Measurements of the RI of tissue with OCT were originally proposed by Tearney et al. (1995), demonstrating two methods to measure the RI of the layers of human skin \cite{Tearney:95}. The first method, the path-length matching method, extracted the RI by comparing the optical and geometrical thicknesses. The second method used OCT focus tracking, which measured the ratio between the optical path length measured by OCT and the resulting focus shift from moving the focus of the objective lens through the sample. Since then, applications based on either of these methods have been used to measure the RI of human crystalline lenses \cite{Uhlhorn20082732}, \textit{in-vitro} human teeth \cite{Mengteeth}, and the mouse crystalline lens \cite{Chakraborty201462}. The path-length matching method is a simple non-invasive method which presents many advantages such as its rapid measurements, increased accuracy and easy operation, in comparison to the OCT focus-tracking method \cite{Mengteeth}. \\

This study investigates the RIs of multicellular spheroids and the effects of hypoxia on multicellular spheroids of the cell line HCT116, a human colon cancer cell line. The motivation of this study is twofold. First, the spheroid RI was measured with OCT utilising the path-length matching method. This enabled the selection of an appropriate RI matched immersion liquid for confocal microscopy to reduce RI mismatch between the spheroid and immersion medium,  so that greater depth of imaging through the spheroid can be achieved as less light scatter will occur. Two immersion media were chosen based on the results of the OCT RI measurements. Second, we investigated whether OCT could be used to monitor RI and provide quantitative information on spheroid growth and cell death. \\ 

\section*{Methods}

\subsection*{Sample preparation} 

The HCT116 colon cancer cells (American type tissue culture collection (ATCC), VA, USA) were grown in a standard incubator at 37{\textdegree}C with 20\% oxygen and 5\% carbon dioxide as monolayer cultures in T75 flasks containing $\alpha$ Minimum Essential Medium (MEM) (Sigma‐Aldrich, USA) with 5\% foetal calf serum (FCS). These cells were dissociated with trypsin and spheroids were cultured by seeding 1000 cells per well in round bottom 96 well plates (Corning, USA),  consisting of surfaces that allow minimal cell attachment, encouraging the cells to form a spheroid. Every second day, the spheroids were replenished with 50\% new growth medium containing 10\% FCS and 1\% penicillin with streptomycin (P/S). \\

The pilot study (Experiment 1) used SiHa cells (ATCC, VA, USA) prepared in the same way as above. The spheroids were fixed and imaged with OCT after immersion in phosphate buffered saline (PBS). Thereafter, three OCT experiments were conducted with fixed HCT116 spheroids which were immersed in ethanol (Experiment 2) or PBS (ACSRC, The University of Auckland, NZ) (Experiments 3 and 4) for 24 hours prior to OCT imaging. The fixation procedure included aspirating the media and immersing the spheroid in 10\% neutral buffered formalin (NBF) for 24 hours at 4°C, then retained in 70\% histology ethanol. To compare the impact of seeding density  on growth, spheroids were also prepared at a lower seeding density of 500 cells per well (experiment 4) and grown in normoxic conditions until day 4 followed by fixation as above.\\

Brightfield images of live spheroids were acquired daily using the Molecular Devices ImageXpress Micro XL (Molecular Devices USA), prior to fixation. These images were used to measure growth characteristics such as volume and diameter using a macro (GB Spheroid Macro), developed by Gib Bogle (Auckland Bioengineering Institute, University of Auckland), for the open source software, ImageJ (NIH, USA) \cite{Mao2}.

\subsection*{OCT image acquisition and analysis}

To determine the spheroid RIs, a custom-built spectral domain OCT (SD-OCT) system \cite{lippok2012dispersion} acquired two-dimensional images (B-scans) of the sample on a flat reflecting surface. To avoid degradation of the samples, images were taken within five minutes of removal from solution. The images of the spheroids were taken where the largest cross-sectional area could be visualised. The SD-OCT employs a 3 x 1 super-luminescent diode (Superlum, Ireland) with a central wavelength, $\lambda_{0}$, of 840~nm, and bandwidth $\Delta \lambda$ of $\approx$100~nm, thus yielding a theoretical axial resolution in air of 3.1 $\mathrm{\mu m}$. The signal-to-noise ratio of the system was measured to be approximately 95~dB. The coupled lateral resolution, $\Delta x$, and the depth of focus, $b$, was measured to be 21~$\mathrm{\mu m}$ and 820~$\mu$m, respectively. The power in the sample arm was approximately 6~$\mathrm{{\mu}W}$, which did not induce any visible damage upon the sample. \\

\begin{figure}[ht]
  \centering
\includegraphics[width=0.8\textwidth]{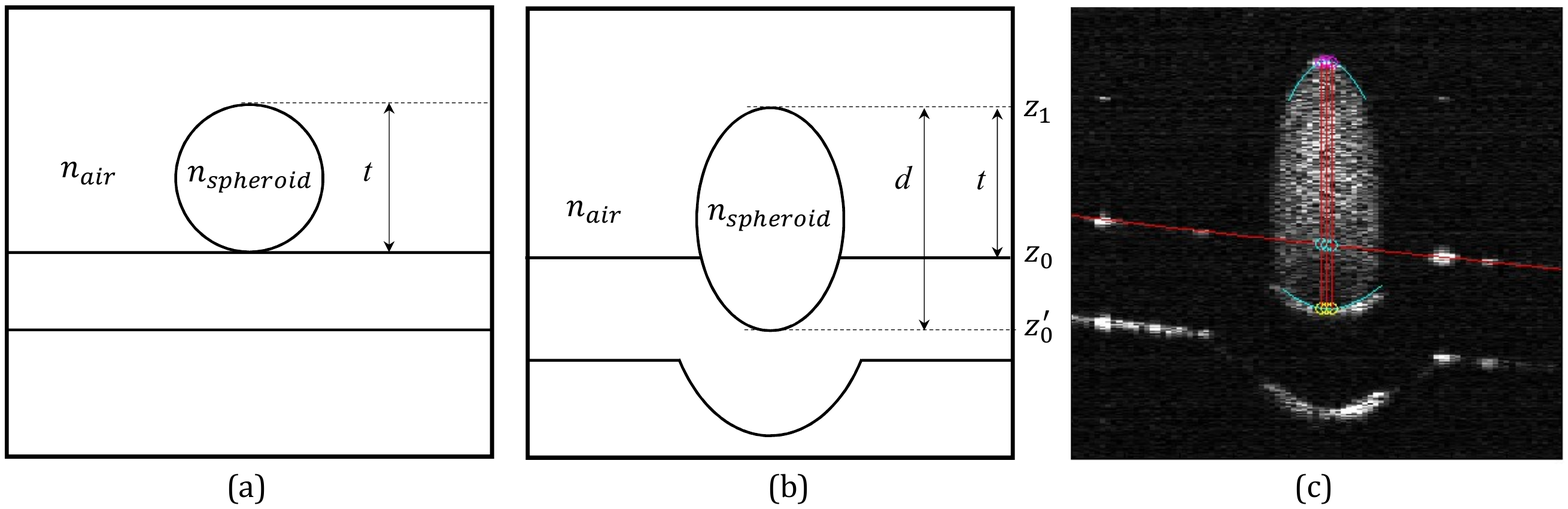}	
\caption{Schematic representation of the spheroid thicknesses as seen physically (a) and by OCT (b). In image (c), the OCT B-scan of the spheroid on petri dish show the fitted lines used to determine the geometrical thickness ($t$) and optical thickness ($d$).}
	\label{fig:schematic}
\end{figure}

OCT directly measures the group delay of light through a sample via coherence gating. A single depth scan (A-scan) in OCT represents the depth-dependent intensity of reflections or back-scattering along the beam path. The optical thickness of a sample, is measured as the distance, $d$, between reflection peaks in an A-scan, also expressed as the product of the group RI, $n_{g}$, and the geometrical thickness, $t$, of the medium.  Thus, the length of a sample depicted in an OCT image (for example, the vertical red lines in Figure~\ref{fig:schematic}) is not representative of the geometrical thickness, $t$. By making an approximation of the sample’s geometrical thickness directly from the OCT image, the group RI of the samples can be determined. The group RI, $n_{g}$, is expressed in terms of the wavelength dependence of the phase RI, $n_{p}$, as:
\begin{equation}
\centering
n_{g} = n_{p} - \lambda_{0}\left( \frac{dn_{p}}{d\lambda}\right)_{\lambda_{0}} ,
\end{equation}
where $\lambda_{0}$ is the central wavelength of the OCT light source. The $\frac{dn_{p}}{d\lambda}$ term is typically disregarded in practice since it is small and negligible \cite{Tearney:95, Mengteeth}. Thus, we assume the group RI of the sample to be similar to the phase index. \\

The multicellular spheroid RI is calculated by taking measurements of the optical thickness and geometrical thickness for each individual spheroid from a single OCT B-scan (Figure~\ref{fig:schematic}). The approximated geometrical thickness is taken as the distance from the top of the spheroid, $z_{1}$, to the upper reflection of the petri dish, $z_{0}$, while the optical thickness is taken as the reflections from the top of the spheroid, $z_{1}$, to the bottom of the spheroid, $z_{0}'$. The ratio of the optical thickness, $d$, to the geometrical thickness, $t$, yields the RI (Equation~\ref{eqn:ri2}):

\begin{equation}
\centering
n_{s}= \frac{z_{1}-z'_{0}}{z_{1}-z_{0}}
\label{eqn:ri2}
\end{equation}

OCT B-scans were taken of spheroids fixed (as described above) at days 4, 5, 6, 7 of their growth cycles for the pilot study (SiHa, Experiment 1) and three different experiments (HCT116, Experiment 2 to 4). At least five OCT B-scans were acquired for each spheroid. All five images from each spheroid were analysed twice with each analysis yielding three RI measurements per image. As a result, each RI value for one spheroid is an average of at least 30 measurements.  The measured RIs of the fixed spheroids were compared to study the effect of growth time and cell seeding number.  \\

The OCT data was corrected for background noise and the analysis is based on the assumption that the multicellular spheroid is spherical in geometry. A MATLAB (Mathworks, USA) code, was created to semi-automate the calculation of the spheroid RI. Figure~\ref{fig:schematic}(c) displays the second-order polynomials which were fitted to the first reflection of the petri dish surface (red line), and the top and bottom surfaces of the spheroid (blue line). The fittings enabled the measurements of the optical thickness and the geometrical thicknesses. The above procedure is conducted for each experiment replicate, meaning that we have at least 3 RI values per day per experiment, each being the average of at least 30 measurements. \\

\subsection*{Confocal microscopy image acquisition and analysis}

\subsubsection*{Fluorescence staining of fixed spheroids}

Fixed spheroids were permeabilised in PBStt (600~mL MilliQ water, 6~g Tween 20 (Global Science), 200~ml 10x PBS, 0.08~g Thimerosal (Sigma-Aldrich, USA), 0.25~g Sodium Azide (Sigma-Aldrich, USA)) for 60 minutes at 4{\textdegree}C. Following the removal of PBStt, spheroids were washed with PBS and stained with either 8~$\mu$M Hoechst 33342 (1~mg/mL) or 8~$\mu$M  propidium iodide (PI; 1 mg/mL), both prepared in PBS for a period of 24 hours at 4{\textdegree}C. Stained spheroids were kept in PBS at 4{\textdegree}C until required for imaging, and light exposure was minimised.\\

\subsubsection*{Image acquisition}

In the initial experiments, spheroids were transferred to ibidi 8 well $\mu$-chamber slides (Ibidi, Germany) for confocal imaging as these have a bottom surface that is equivalent to a coverslip in both thickness and optical clarity. Each spheroid was placed in a separate well.  The spheroids were kept hydrated in PBS (n $\approx$ 1.33) and imaged using a Zeiss LSM 710 inverted confocal microscope (Carl Zeiss AG, Germany) with a 10x/0.45 NA Plan Apochromat dry objective lens.\\

\begin{figure}[ht]
  \centering
\includegraphics[width=0.6\textwidth]{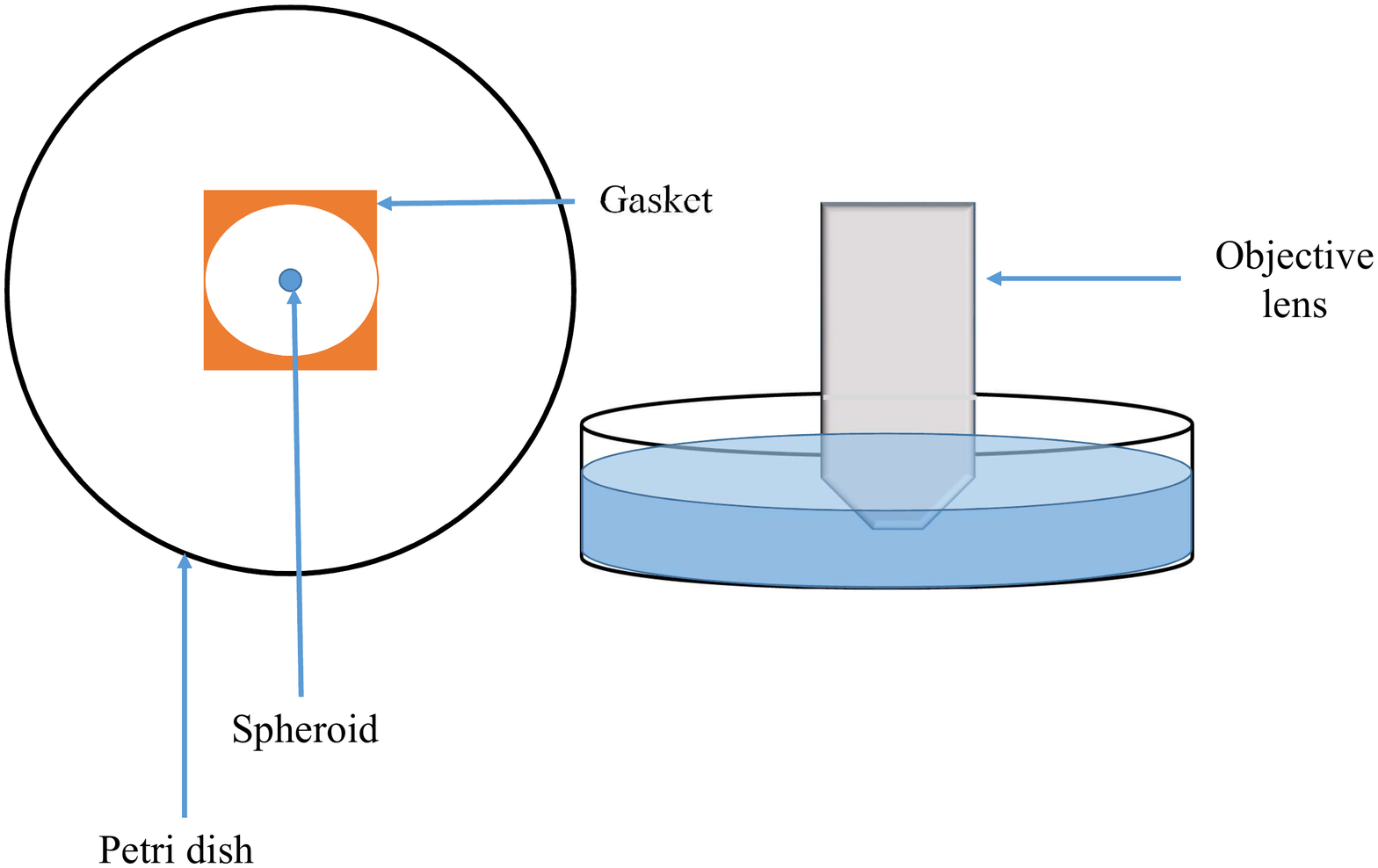}	
\caption{For immersion confocal imaging on the Olympus FV1000 upright confocal microscope, fixed spheroids were placed in a 50~mm petri-dish, with a droplet of mountant and sealed off by a 1~mm gasket in order to keep the spheroid stationary. The petri dish was filled with enough immersion medium to accommodate the 10X/0.6NA XMLPV10X immersion objective lens.}
	\label{fig:confocal}
\end{figure}

Following the pilot OCT study, improvements in the confocal imaging protocol were made based on the RI measurements. Two types of immersion media, which are also optical clearing agents, were identified as being appropriate for testing: glycerol (n = 1.473;  Merck, New Zealand) and ScaleView-A2 (n = 1.380; Olympus Corporation, Japan). An Olympus FV1000 upright confocal  microscope (Olympus Corporation, Japan) with a specialised immersion lens (XLPLN 10x SVMP/0.6 NA) was used to image the spheroids in glycerol or ScaleView-A2. PBS (n = 1.333) immersion was also trialled. The objective lens has an adjustable correction ring for different RIs. For confocal microscopy, the spheroid was placed in a 90~mm petri dish and sealed with a CoverWell imaging chamber gasket with built-in coverslip (Life Technologies Corporation, USA) (Figure~\ref{fig:confocal}) so that the spheroid would not be affected by movement of the objective lens, keeping the spheroid stationary. Sufficient immersion medium was then added to accommodate the objective lens. The objective lens used has a large diameter and 8~mm working distance thus requires a large volume of immersion medium to be used. A z-series, with a step size of 7~$\mu$m, encompassing all of the visible staining, was acquired for each spheroid.  Excitation/emission wavelengths were 405/411 - 461~nm (Hoechst 33342) and 561/566 – 718~nm (PI).\\

\subsubsection*{Image analysis}
To determine the performance of each immersion media, analysis was performed for each z-stack from representative day 5 spheroids immersed in glycerol, ScaleView-A2 and PBS using ImageJ. A region of interest that represented the area of the whole spheroid was selected. A specialised plugin for ImageJ developed by Maske~\cite{MaxGitHub} was used to derive the raw integrated density, which describes the strength of the signal and is determined by the sum of the values of the pixel within the specific area. The depth of imaging was determined at the point where no fluorescence signal above background was detected.\\

\subsection*{Detection of necrosis in live spheroids}  

Necrosis in live spheroids was defined based on a protocol used by Zhang et al. (2016) \cite{zhang2016optimization}. Spheroids were seeded at 1000 cells per well. Individual spheroids were removed on days 4, 5, 6, 7 and 9 with 100~$\mu$L of media and placed into eppendorf tubes with 6~$\mu$L of PI (1~mg/mL)(8$\mathrm{\mu}$M in PBS). Tubes were incubated at 4{\textdegree}C for 90 minutes, after which spheroids were replaced back into wells with fresh culture media, $\mathrm{\alpha}$-MEM + 10\%~FCS + 1\%~P/S. Live spheroids were imaged on the ImageXpress, using the Triple 2 (TRITC) filter set with excitation 549/15~nm and emission 580-648~nm wavelengths.\\

\section*{Results}

\subsection*{RI quantification of spheroids}

\begin{figure}[h!]
  \centering
\includegraphics[width=0.7\textwidth]{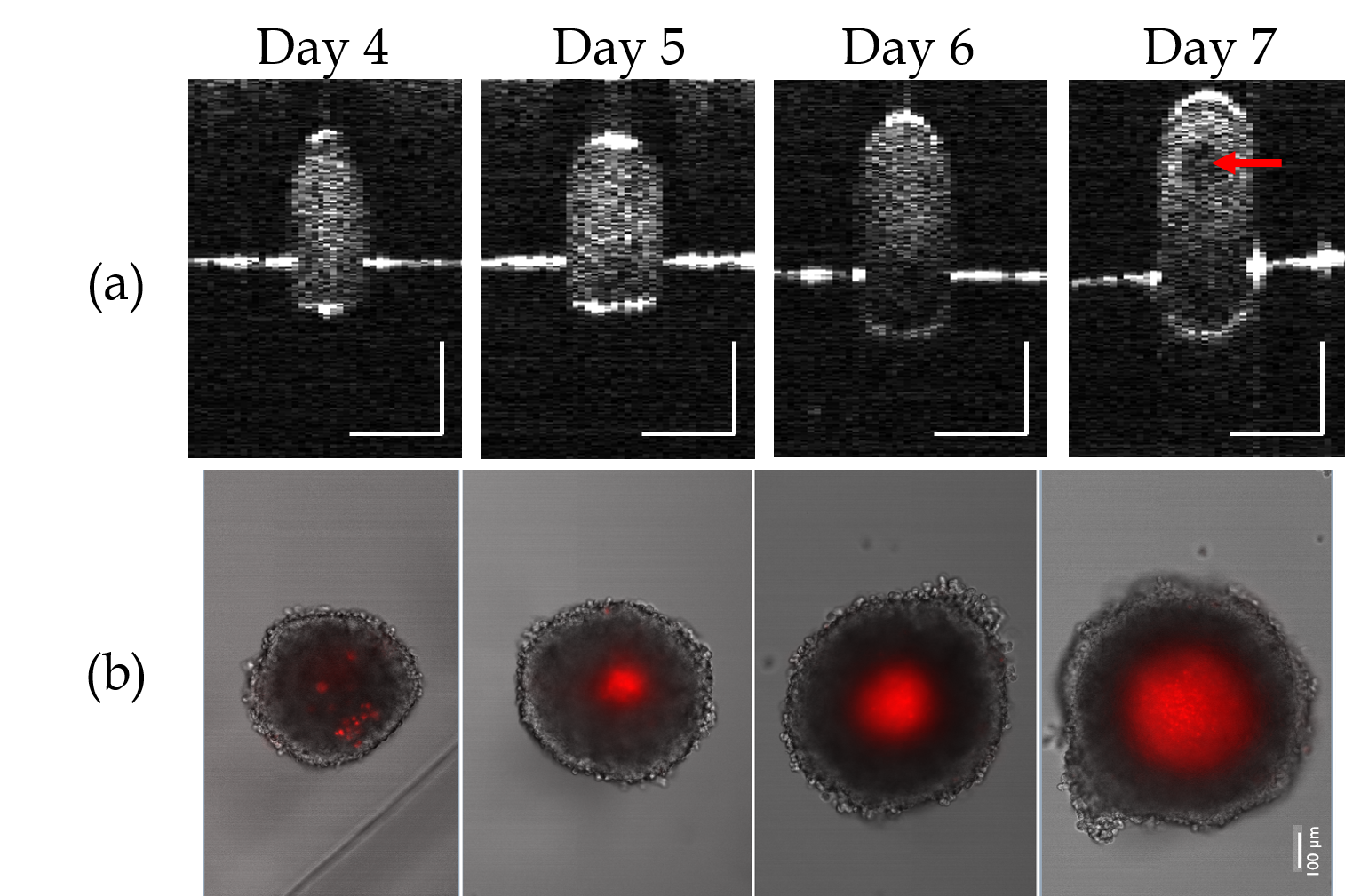}	
\caption{(a) Evolution of spheroid structure as imaged by OCT. Cross-sectional OCT images showing the growth of the spheroids with day and the presence of necrosis (red arrow) at day 7. Scale bars on OCT images represent 500~$\mu$m.(b) Brightfield images of spheroids merged with widefield fluorescence imaging of the propidium iodide identifying dead cells in the necrotic core.}
	\label{fig:OCTspheroid}
\end{figure}

Representative spheroids visualised with OCT with their corresponding merged brightfield and fluorescence microscope images are shown in Figure~\ref{fig:OCTspheroid} a and b, respectively. The OCT images display the full depth of the fixed spheroid showing spherical morphology. OCT images of spheroids greater than day 5 display clear dark regions near the centre of the spheroids (Day 7 in Figure~\ref{fig:OCTspheroid}), corresponding to regions of necrosis, due to the higher OCT signal attenuation. However, near the bottom of the spheroid, the lack of OCT signal intensity present is not representative of necrosis.\\

\begin{figure}[ht]
  \centering
\includegraphics[width=0.8\textwidth]{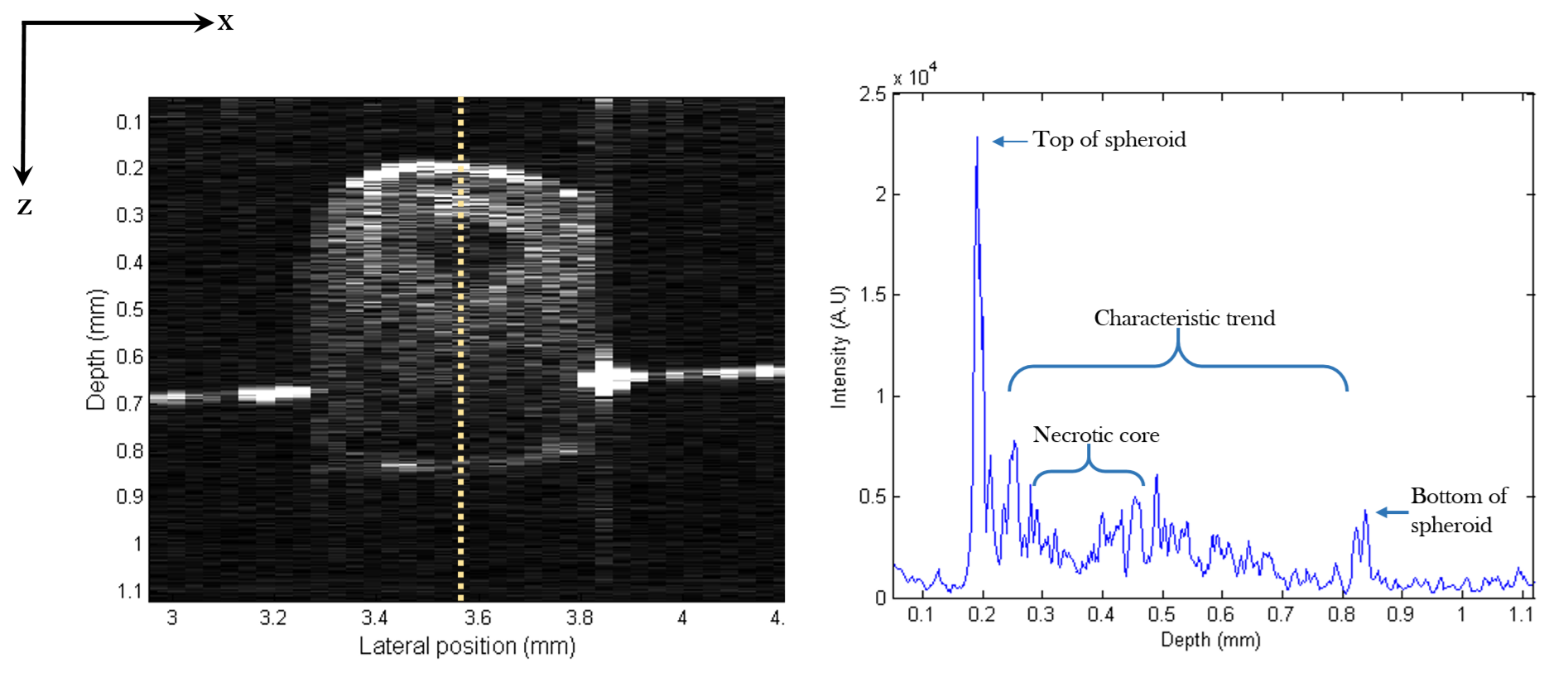}
\caption{(Left) B-scan of HCT116 spheroid at day 7. (Right) Depth scan profile taken at x = 3.56~mm. A dip in the profile between 0.3~mm and 0.5~mm coincides with the dark necrotic region in the B-scan on the left.} 
	\label{fig:depthgraph2}
\end{figure}

Figure~\ref{fig:depthgraph2}, which displays an A-scan passing near the centre of the spheroid, exemplifies the fluctuations in signal with depth as well as the attenuation of the signal with depth, which is a typical characteristic with OCT A-scans, before detection of a major reflection of the bottom of the spheroid as at approximately 0.85~mm. The dip in signal at approximately 0.3~mm and 0.5~mm corresponds to the signal from the necrotic core.  \\

\begin{figure}[h!]  
  \centering
\includegraphics[width=1\textwidth]{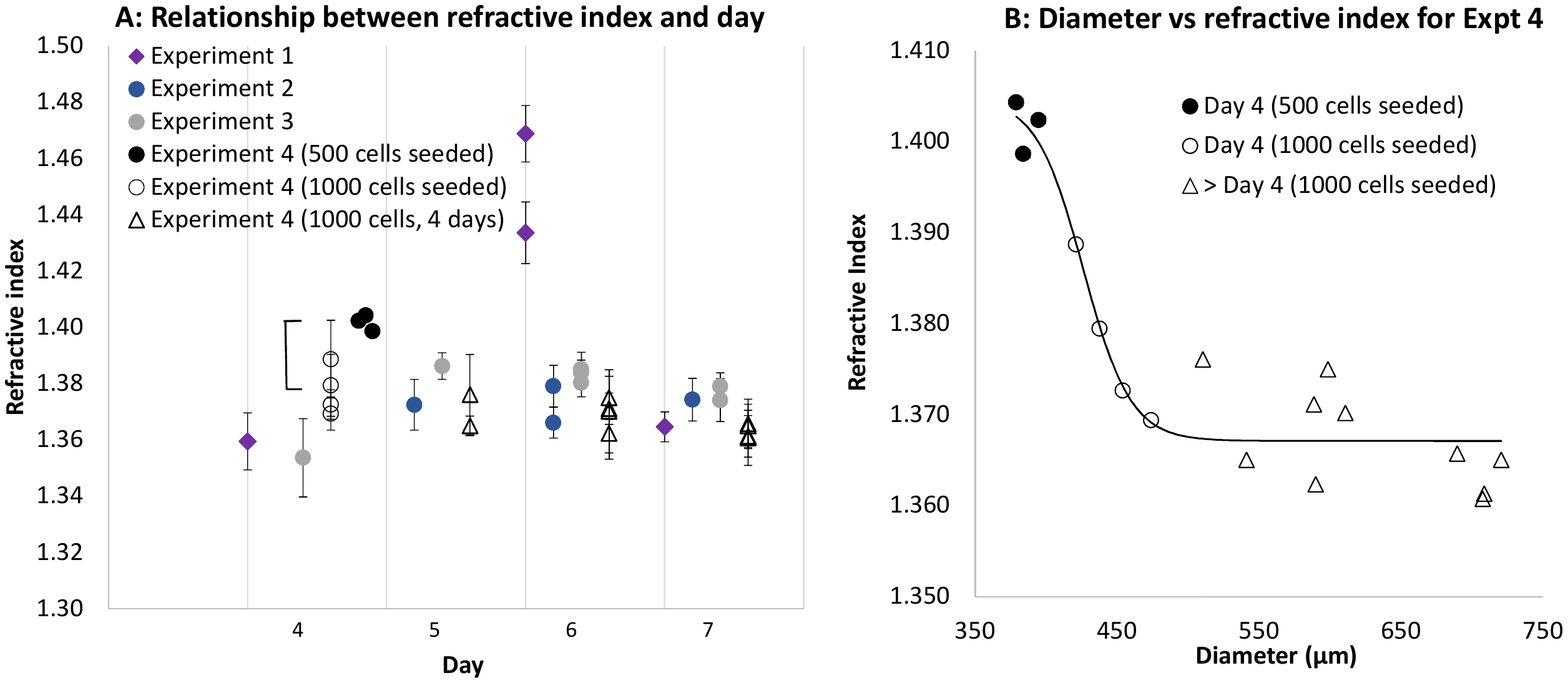}
\caption{(A) Measured RIs (n $\pm$ $\sigma$) for each replicate at each day including all experiments. (B) Dependence of RI on spheroid diameter (measured by live spheroid imaging) for experiment 4}
	\label{fig:RIall}
\end{figure}

The average RI was quantified through the centre of each spheroid (Figure~\ref{fig:RIall}, left panel). The measured RIs for spheroids seeded at 1000 cells per well (n = 1.35 to 1.39) show no significant difference between days (p-value\textgreater 0.05; One-way ANOVA), despite the onset of necrosis visible at day 7 by OCT. The RI values of day 4 spheroids seeded at 500 cells per well (n~$\approx$ 1.39) are significantly higher than spheroids from the same experiment seeded at 1000 cells per well (n~$\approx$~1.37) (p-value\textless 0.05; One-way ANOVA). Figure~\ref{fig:RIall}, right panel shows the decreasing RI with increasing spheroid size as measured by live spheroid imaging for experiment 4.\\

\subsection*{Immersion media matching}

\begin{figure}[h!]
  \centering
\includegraphics[width=1\textwidth]{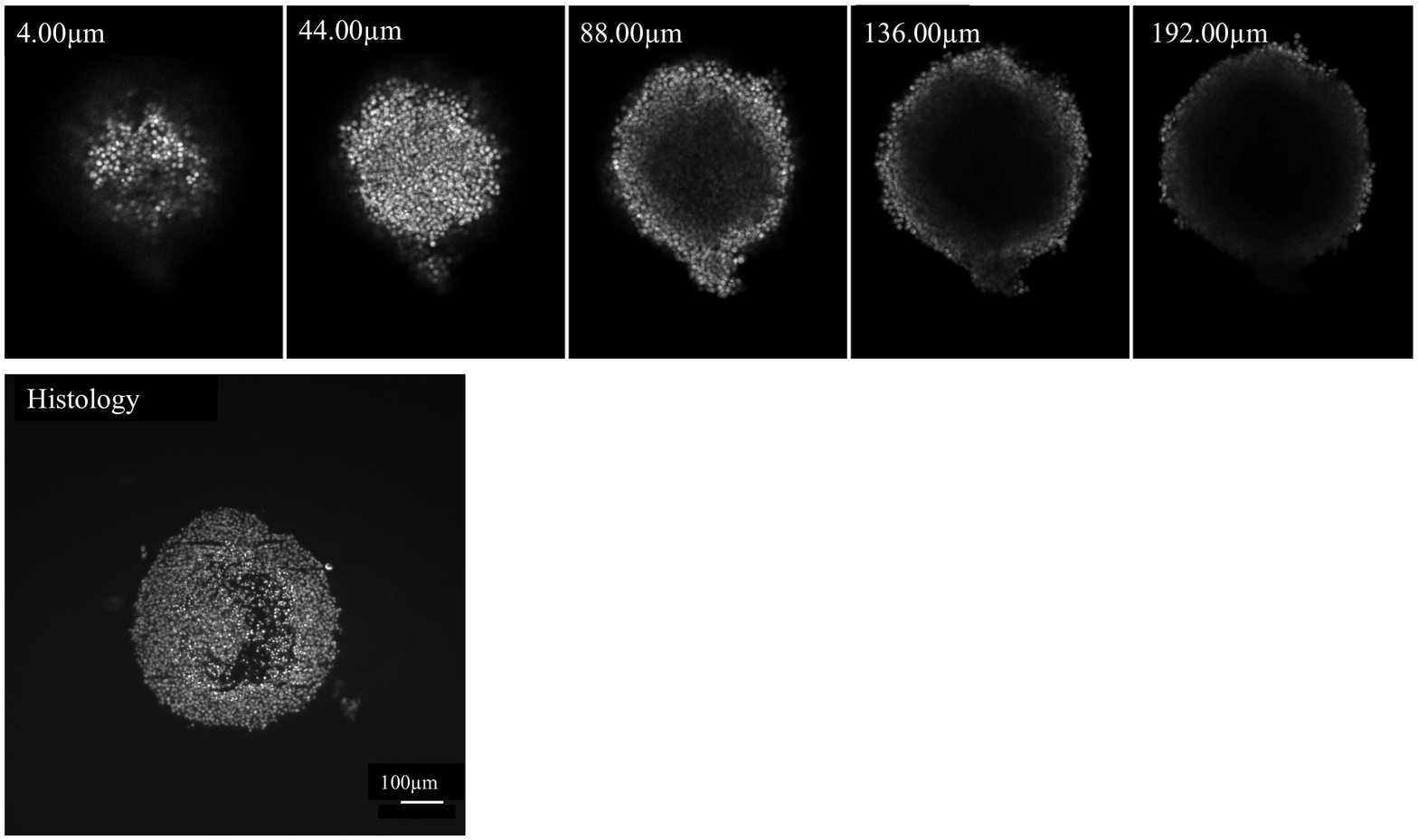}	
\caption{Upper images: SiHa spheroids were cultured with a seeding density of 1000 cells per well and were fixed in 10$\%$ NBF at 4{\textdegree}C for 24 hours, followed by transfer into 70$\%$ histology ethanol. Fixed spheroids were exposed to Hoechst 33342 (8~$\mu$M) overnight. An imaging depth of 192~$\mu$m (step size = 4~$\mu$m) was achieved for a day 6 SiHa spheroid stained with Hoechst 33342 for 24 hours using the Zeiss LSM 710 inverted confocal microscope, 10x/0.45NA Plan Apochromat dry objective lens, and 405~nm diode. The images are from a z-series of a representative day 6 SiHa spheroid showing the early loss of signal, particularly in the central region. The z position of each image is displayed in the corner.  Lower image:  Image of a central transverse histological section from a similar spheroid stained with Hoechst 33342 for 24 hours before sectioning, showing complete dye penetration and a thick viable rim which is not visible with confocal microscopy.}
	\label{fig:dye}
\end{figure}

Initially when PBS (n $\approx$ 1.33) was used as an immersion medium in conjunction with confocal imaging on a Zeiss 710 LSM and a dry lens, the confocal microscopy setup produced three-dimensional images of the spheroids with very limited imaging depth. Figure~\ref{fig:dye} shows that at a depth of 192~$\mu$m, there was very little signal apparent. As demonstrated by the z-series shown in Figure~\ref{fig:dye}, a central dark area devoid of staining was observed. However, histological analysis of serial sections confirmed that 24 hour incubation with Hoechst 33342 stained the entire spheroid. A viable rim was identified which was much thicker than demonstrated by confocal microscopy, indicating an issue with light penetration, rather than lack of staining in the centre.\\
 
The RI measurements aided in improving the depth of imaging of the spheroids with confocal microscopy. In particular, the day 4 and 5 spheroids, when less or no necrosis was visualised, as demonstrated by PI staining (Figure~\ref{fig:OCTspheroid}), were valuable for identifying the two immersion media selected for trial. To improve the imaging conditions, the knowledge of the spheroid RI obtained from the OCT measurements allowed us to select appropriate immersion media that minimised the mismatch in RI. Two immersion media, glycerol (n = 1.473) and ScaleView-A2 (n = 1.380), matching the spheroids RIs were trialled with confocal microscopy. ScaleView-A2 was chosen as the results from the OCT measurements indicated a RI between 1.35 and 1.39. However, the pilot OCT study with SiHA cells reported higher RIs at day 6 of 1.43 and 1.46, which indicate that glycerol, despite its higher RI than the HCT116 spheroid RI, might be an appropriate immersion medium. Moreover, glycerol is a commonly used optical clearing agent and therefore provides a good comparison with previous studies. The protocol was further improved with the use of a specialised immersion lens. Both glycerol and ScaleView-A2, when employed with the specialised immersion lens, enabled better imaging depth than the initial immersion liquid PBS (Table~\ref{tab:depth}). \\

\begin{figure}[ht]
  \centering
\includegraphics[width=1\textwidth]{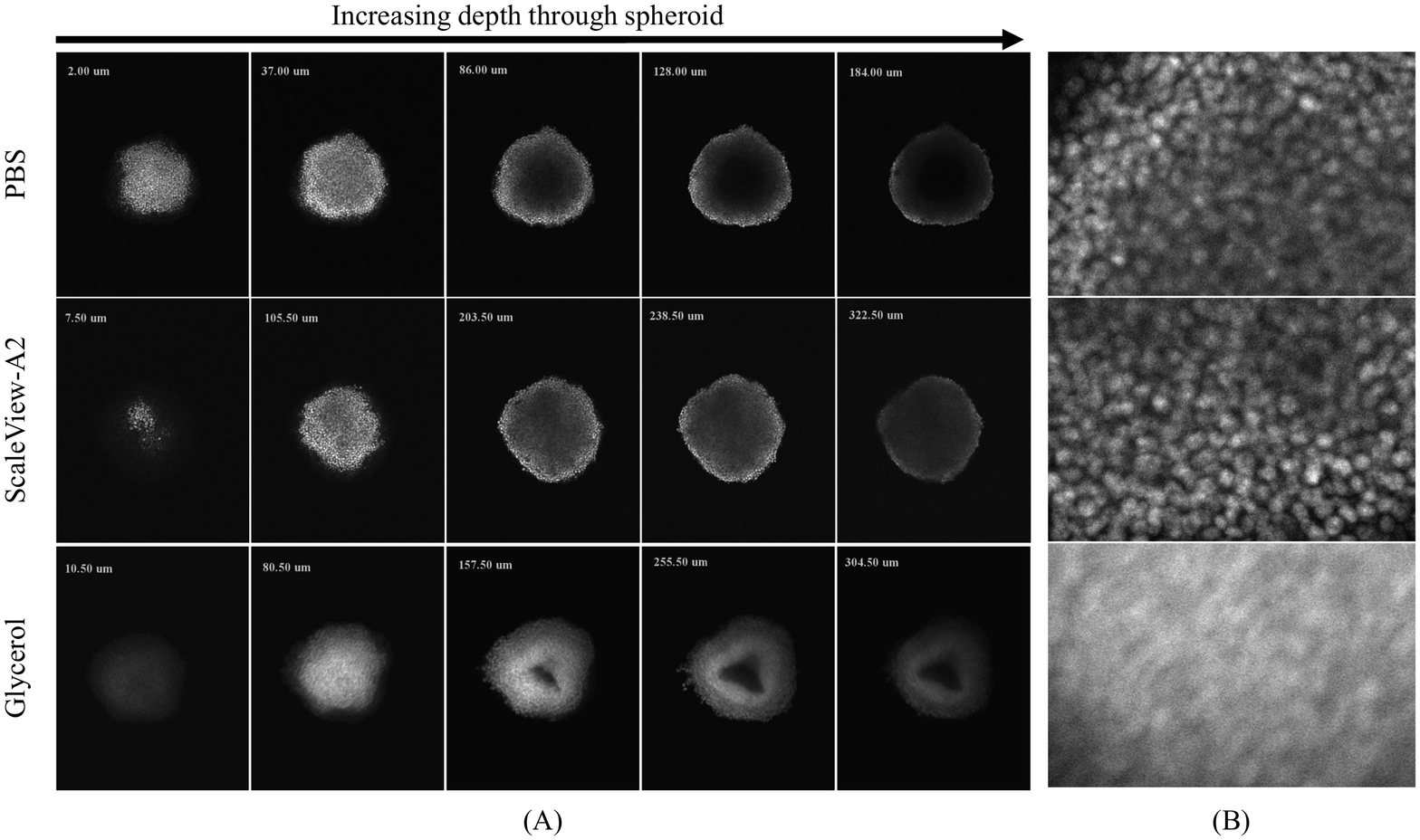}	
\caption{(A) Single confocal optical sections from a z-series of Hoechst 33342 stained HCT116 spheroids immersed in PBS, ScaleView-A2 or glycerol. The z position of each image is displayed in the corner.  HCT116 spheroids, seeded at a density of 1000 cells per well, were fixed on day 5 in 10\% NBF. They were stained with Hoechst 33342 (8~$\mu$M) for a period of 24 hours at 4{\textdegree}C. Spheroids were then kept in either ScaleView-A2 or glycerol for a period of 24 hours prior to being imaged on the Olympus FV1000 confocal microscope, using the immersion objective (XLPN10XSVMP), 10x/0.6NA. (B) Close-up of PBS at 128~$\mu$m, ScaleView at 238.50~$\mu$m and glycerol at 255.50~$\mu$m }
	\label{fig:dye2}
\end{figure}

\begin{table}[htb]
\centering
\begin{tabular}{c|c|c|c|c}
\multicolumn{5}{c}{Depth achieved ($\mu$m)}   
 \\
 \hline
\multirow{2}{*}{} & \multicolumn{2}{l|}{Propidium Iodide}  & \multicolumn{2}{l}{Hoechst 33342} \\ \hline
                  & Mean          & SD (N)         & Mean         & SD (N)         \\ \hline
PBS (n$\approx$1.33)&220            &25 (8)         &188     &  20 (6)         \\
Glycerol (n=1.473)  &440        &21 (2)      &520      &80 (7)           \\
ScaleView A2 (n=1.380)&444       &34 (5)           &424           &0 (1)          
\end{tabular}
\caption{Depth of z-stacks, indicative of imaging depth, achieved in confocal microscopy with different immersion liquids, PBS, glycerol, and ScaleView-A2. Spheroids were seeded at 1000 cells per well and grown for 4 or 5 days. The depth varied with dyes: Propidium Iodide and Hoechst 33342. Microscopy was performed on the Olympus FV1000 confocal microscope.} 
\label{tab:depth}

\end{table}

When spheroids were imaged while immersed in PBS, the depth of imaging was very restricted as demonstrated by the early onset of signal attenuation (Figure~\ref{fig:dye2}, top) with a total imaging depth of only 180-220~$\mu$m. A central dark area devoid of staining was observed although histological analysis had previously confirmed that 24 hour incubation with Hoechst 33342 stained the entire spheroid and that there were cells present in the centre.

\begin{figure}[ht]
  \centering
\includegraphics[width=0.7\textwidth]{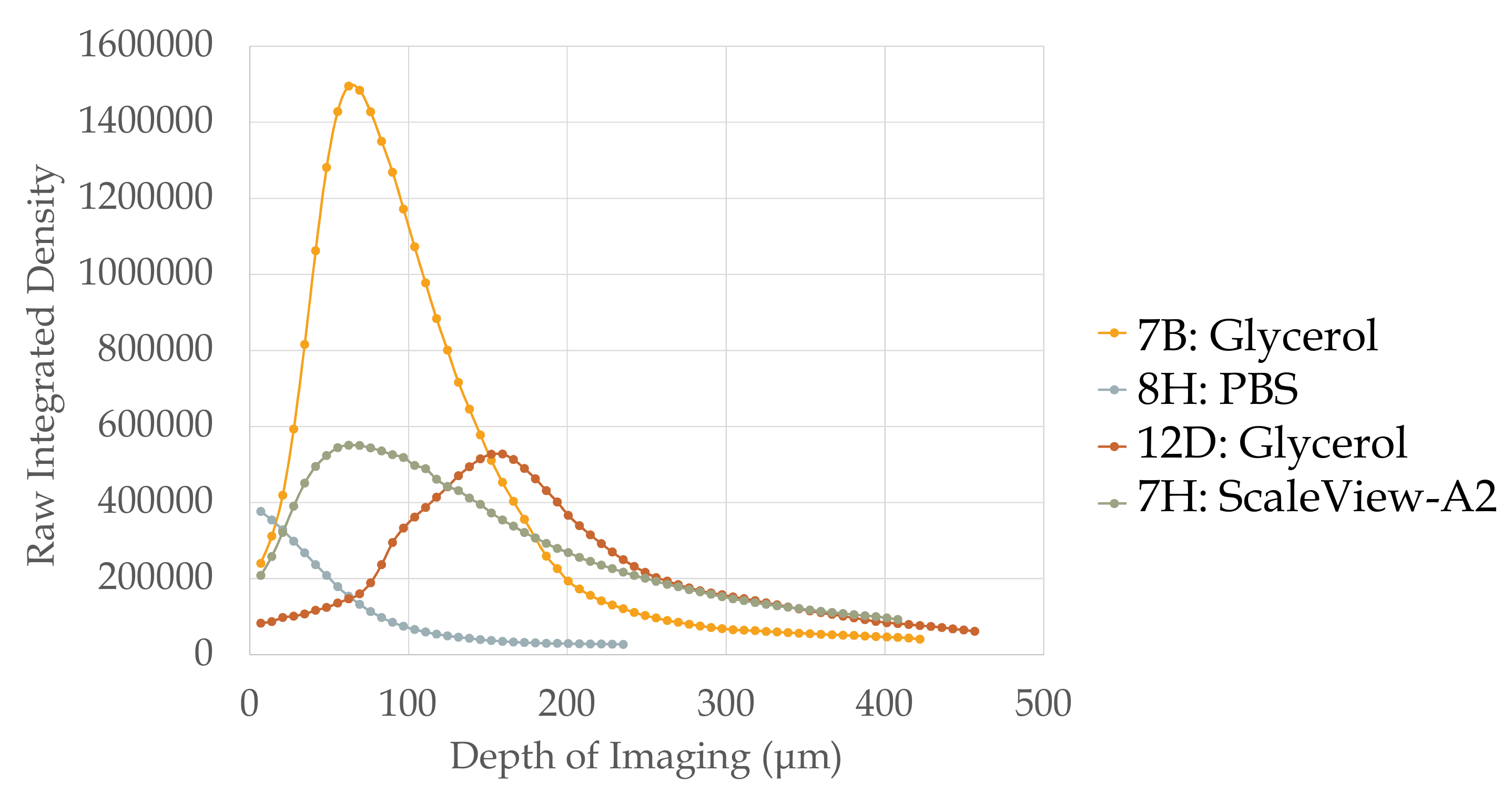}
\caption{Graph showing mean signal through a z-series for individual spheroids immersed in different media. HCT116 spheroids, seeded at a density of 1000 cells per well were fixed on day 5 in 10\% NBF. They were stained with Hoechst 33342 (8~$\mu$M) for a period of 24~hours at 4{\textdegree}C. Spheroids (excluding those in PBS) were then kept in either ScaleView-A2 or glycerol for a period of 24 hours prior to being imaged on the Olympus FV1000 confocal microscope, using the immersion objective (XLPN10XSVMP), 10x/0.6NA. Each point represents a measurement in the z-series.} 
	\label{fig:depthgraph}
\end{figure}

Glycerol enabled a greater penetration depth of at least 413~$\mu$m (Table~\ref{tab:depth}). However, in comparison to ScaleView-A2 and PBS, the structural integrity was compromised and the individual nuclei were less distinct. A central dark area was evident when glycerol immersion was employed and for both PBS and glycerol immersion, the profile of the Hoechst 33342 stain was not uniform, making it initially difficult to discern whether the central dark area was due to necrosis or limited penetration of the stain into the spheroid (Figure~\ref{fig:dye2}, middle). In terms of signal intensity or raw integrated density, glycerol showed inconsistency with a peak and sharp decrease, in addition to two different curves (Figure~\ref{fig:depthgraph}).

ScaleView-A2 provided a better RI match with the HCT116 spheroid experiments. Although the penetration depth (at least 406~$\mu$m) achieved by ScaleView-A2 was not greater than that achieved with glycerol, the individual nuclei were better resolved (Figure~\ref{fig:dye2}). Moreover, the visualisation of the staining was more uniform with a more gradual decrease in signal intensity throughout the spheroid than with glycerol, and the structural integrity of the spheroid was not compromised. 

\subsection*{Diameter measurements}
\begin{figure}[h!]
  \centering
\includegraphics[width=0.8\textwidth]{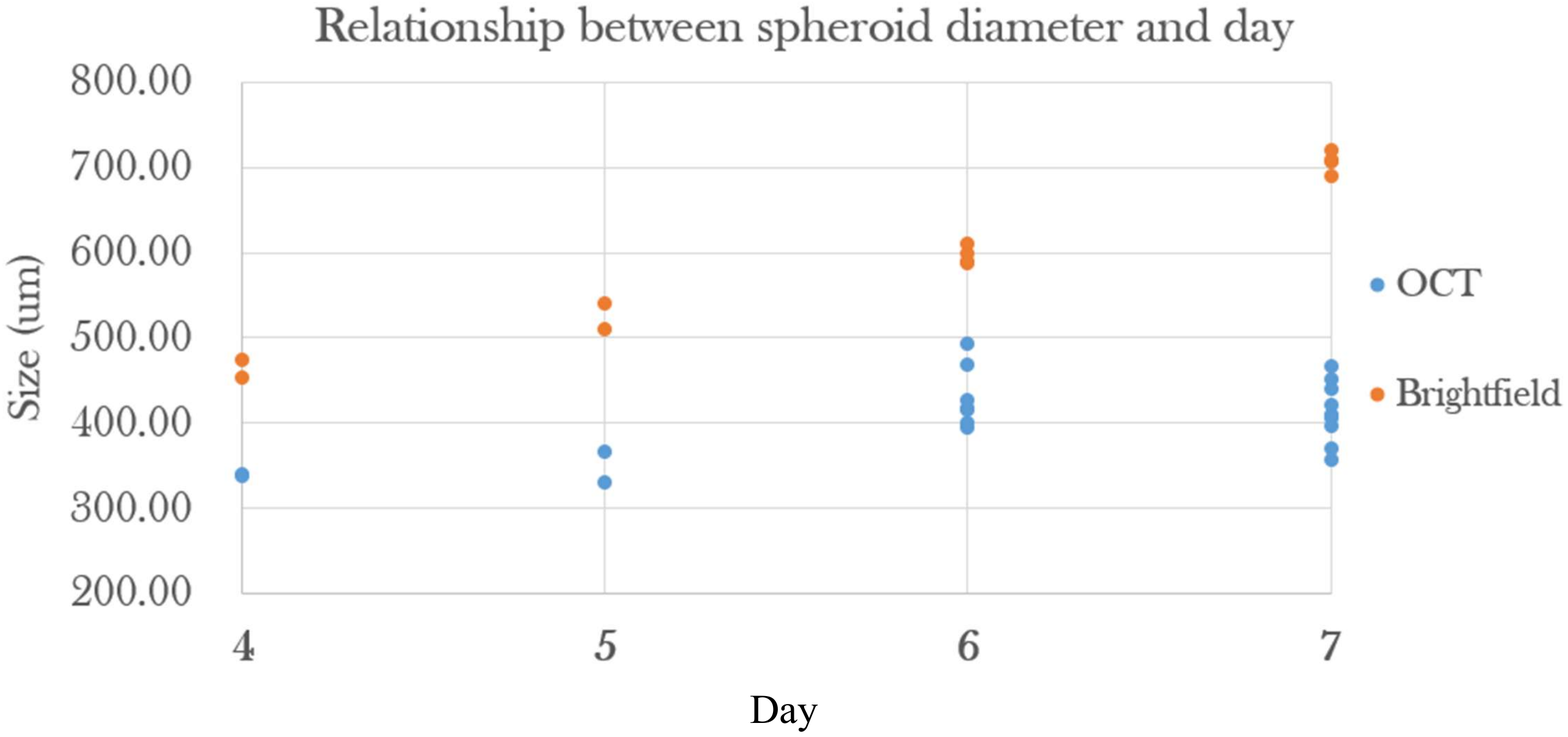}
\caption{Growth of the HCT116 spheroids used in this study as a function of time, as measured from brightfield images (spheroid diameter) of live spheroids using the ImageXpress. For comparison, measurements on the same spheroids made by OCT (geometrical thickness, $t$) are shown.} 
	\label{fig:size}
\end{figure}

Measurements of spheroid diameter by brightfield microscopy allowed comparison with measurements by OCT (geometrical thickness, $t$). The trends are similar with lower values of diameter obtained by OCT (Figure~\ref{fig:size}), presumably due to shrinkage during fixation.

\section*{Discussion}

The images and measurements taken with OCT open up many possibilities for the study of multicellular spheroids in cancer research. It is important to note that OCT allows us to visualise the full depth of the spheroids in comparison to confocal microscopy. With OCT, we measure the RI of a multicellular structure averaged along the path of light through the centre of the spheroid. The measurement is a result of a combination of different cells, possibly at different stages of growth. At a microscopic scale, the RI of a single cell is related to the concentration of cellular components within it. At the subcellular level, a cell can be considered as an entity with a spatial RI distribution \cite{LiuRefractive} contributed by the organelles. The RI of a cell is a result of many microscopic fluctuations of the RI within the cell \cite{martelli2004effect}. When extending this to a multicellular spheroid as a cluster of single cells, we can assume a model with normal (viable) cells near the periphery, necrotic cells (dead) at the centre, and apoptotic cells (dead) distributed throughout. As a result, RI values of the spheroids reflect the typical values of cellular component RIs, which range from n = 1.355 for the nucleus to n = 1.600 for lysosomes as summarised by Liu et al. (2016) \cite{LiuRefractive}.\\

Liu et al. (2016) also mentions that the RI of a cell can be used to monitor the stage of the individual cell growth cycle \cite{LiuRefractive}, as the amount of intracellular components changes at each phase of growth. For example, there is an increase in RI of the cell when the DNA content within the nucleus doubles, which coincides with the transition of the cell from earlier cell growth stages of G1/S to later stages of G2/M \cite{BistaRI2011}. However, many human cancer cells display atypical cell cycles and increase in cell proliferation \cite{LiuRefractive} which may contribute to the higher RI. Previous studies have measured the RI of normal and cancer cells and have confirmed RIs of 1.353 for normal cells and a range from 1.370 to 1.400 for cancer cells \cite{liang2007determining, Choi10}. In comparison to normal cells, cancerous cells contain a larger concentration of protein (n = 1.50 - 1.58) in the nucleus and thus display a greater RI to normal cells \cite{liang2007determining}. The current study confirmed that changes in RI were evident throughout the spheroid. With further analysis such as image segmentation and curve fitting, the RI of the individual components of the spheroids can be measured.\\ 

The difference in the range of spheroid RIs between the pilot study SiHa and the HCT116 cells may be attributed to the fact that the characteristics of multicellular spheroids vary with the origin and type of cell line used \cite{Tannock878, Mao1}. However, one set of measurements of the SiHA spheroids was not sufficient to confirm the difference. Over days 4 to 7 of spheroid growth, the cells rapidly multiplied and the spheroids grew in size, as expected \cite{Mao1}. However, the spheroid RIs did not show a trend over that short period of time, which indicates the need for a longer longitudinal study. \\

The discrepancy between the diameter measurements obtained by OCT and confocal microscopy is possibly due to spheroid shrinkage during the fixation process. The increased discrepancy at day 7 (Figure~\ref{fig:size}) may be attributed to the larger proportion of necrotic debris present compared to earlier time points which might influence the shrinkage process during fixation.  It could also be corrected by a 3D measurement to ensure the diameter is measured exactly at the centre of the spheroid.  \\

There is a clear size dependence of the RI for HCT116 spheroids. OCT measured a higher RI for smaller spheroids (seeded at 500 cells per well) in comparison to the large spheroids (1000 cells per well) in the same experiment. The higher RI of small spheroids implies that RI variations of cells provides useful quantitative information of the onset of necrosis, apoptosis or growth inhibition due to treatments, and is worthy of further study. \\

Cell death generally occurs via two different pathways, apoptosis and necrosis \cite{vermes1994apoptosis}. Apoptosis describes a natural programmed pathway of cell death, while necrosis is often called "accidental" cell death, as it occurs when cells are exposed to extreme variance in conditions, whether thermal, chemical or anoxic. The different processes influence the intracellular water in different ways, where apoptosis is characterised by water loss, while necrotic cells absorb water \cite{kerr1972apoptosis}. Furthermore, apoptosis occurs at any stage of growth and is not confined to only the centre of the spheroid. An overall higher RI was displayed by the small spheroids in comparison to the larger spheroids, thus reflecting a combination of a change in the individual cell’s composition at the subcellular scale and the proportion of different cells at the multicellular scale. \\

Initially, the use of PBS with a dry objective lens, introduced a mismatch between two major interfaces: air/PBS ($n = 1.00/n \approx 1.33$) and PBS/spheroid ($n \approx 1.33/n = 1.35 - 1.39$) resulting in a very shallow imaging depth. Using an immersion lens in PBS did not greatly improve the imaging depth of imaging, which indicated that there was still an issue with light penetration, probably due to the difference in RI between PBS and the spheroid causing light scattering due to refraction at the interface. With the application of glycerol (n = 1.47) or ScaleView-A2 (n = 1.38) in conjunction with the immersion lens, the mismatch was reduced to one interface: media/spheroid. This likely lessened the spherical aberration with depth compared to the original data set. To reduce the mismatch at the immersion liquid/sample interface, the RI of the cells near the cell/liquid interface were theoretically important as the RI mismatches on a macroscopic scale, e.g., between spheroid and immersion medium, determine the refraction of light. \\

Glycerol appeared to perform better in terms of optical clearing, as a greater depth of imaging was achieved and the spheroid was more transparent, however the distorted shape, less visible nuclear details, and inhomogenous staining proved to be drawbacks. When fixing spheroids for microscopic studies, the damaged cell membrane becomes permeable to dyes, and thus the Hoechst 33242 dye is able to penetrate into the centre of the spheroid. However, the  dye, which stains dead cells and the DNA in cell debris, was not seen in confocal images of the centre of the spheroid, when immersed in glycerol. In contrast, when ScaleView-A2, which matched the spheroid RI better than glycerol, was used homogeneous staining was clearly seen at equivalent depths, which was consistent with histological sections, although the depth of imaging achieved was slightly reduced.\\

We have demonstrated that, after day 5, the core of the spheroid is not cellular, but consists of Hoechst 33342 staining cellular debris \cite{Mao1}. A perfect RI match enables no contrast with transmitted light microscopy with the spheroid appearing invisible. All the mechanisms involved in the improved imaging depth achieved when using Scaleview-A2 are not yet completely understood for this particular tissue type, but is likely to be due to the combination of (1) reduction of RI mismatch and (2) effects of the optical clearing process. These two effects could not be distinguished separately, however the knowledge gained from this study will be applied to future studies of whole spheroid imaging of hypoxia by confocal microscopy. \\

OCT has advantages such as its fast image acquisition time, three-dimensional imaging, full depth visualisation of the spheroids and non-destructive and non-invasive technology.  This study identified improvements that should be made such as the need to remove the spheroid from solution before imaging under OCT and fixation of spheroids. Future work will include imaging live spheroids as recently described \cite{huang2017optical} in wells without fixation. Furthermore, lessening the power of the OCT imaging beam upon the sample (6~$\mu$W in this study) will ensure no damage to the sample while imaging. This can furthermore be resolved by imaging the spheroid while submerged in solution. However, strong reflection from the immersion fluid does influence image quality, so the spheroid must be covered in sufficient fluid, so that the air/fluid interface is outside the coherence gate when imaging. In addition, the ideal imaging system includes imaging the spheroids \textit{in-vitro} i.e. without fixation which could be easily achieved with a handheld probe. Future applications will incorporate both fixed and fresh spheroids. \\

Our results highlight that RI has the potential to be used as a marker for spheroid growth phase. The increase in RI in small spheroids implies that variations in the RI of cells can provide more information on growth conditions. In this study, OCT could not provide sufficient information on the ratio of necrosis to proliferating cells in the spheroid, so direct inferences were not possible. The necrosis visualised in the OCT structural images [Figure~\ref{fig:OCTspheroid}, Day 7 (top)] is due to a change in RI, which demonstrates the importance of light attenuation (Figure~\ref{fig:depthgraph}) as a marker of necrosis. Huang’s study confirmed the potential of necrosis being identified through different attenuation coefficients by necrotic cells within the spheroid. OCT allows the quantitative measurement of the local optical attenuation coefficient, $\mu_{t}$, which is described as the decay in intensity against depth. It would be interesting to see the changes in $\mu_{t}$ by day. It shows the potential of OCT for necrotic core visualisation without dyes (attenuation coefficient). Furthermore, a study by van Leeuwen et al. (2010) demonstrated measurements of the attenuation coefficients in viable, apoptotic and  necrotic cells \cite{van2010apoptosis}. This supplementary information of the spheroid is advantageous when distinguishing between the two cell deaths as there is potential for necrotic core visualisation without dyes using the attenuation coefficient.\\

At a broader level, information on the RI, given by OCT, provides increased knowledge of the characteristics of multicellular spheroids of a particular cell line. These informations may help to improve predictions of drug efficacy, when using the multicellular spheroid model for investigations combined with complementary imaging techniques such as OCT and confocal microscopy. The use of such multi-parametric spheroid studies are expected to have an impact in shortening drug discovery timelines, reducing cost of investment, and bringing new medicines to patients sooner. \\

\section*{Conclusion} 
In conclusion, with a simple custom-built SD-OCT, we successfully conducted RI measurements of large (n = 1.35 - 1.39) and small spheroids (n~$\approx$ 1.39 - 1.41) noninvasively and nondestructively with rapid image acquisition. The variation between the spheroids at each day can be attributed to both the change in cell composition and the proportion of live/dead cells. These results suggest that alterations in RI may be a sensitive indicator of changes resulting from growth or treatment. OCT also allows visualisation of necrosis.  Furthermore, the findings were directly implemented in confocal microscopy to reduce the RI mismatch between the spheroid and the immersion medium. ScaleView-A2 aided in achieving greater imaging depths of the multicellular spheroids under confocal microscopy. This improvement in imaging depth confirmed the utility of our RI measurements, proving the promising outlook of OCT in cancer research.

\bibliography{sample}

\begin{thebibliography}{10}
\expandafter\ifx\csname url\endcsname\relax
  \def\url#1{\texttt{#1}}\fi
\expandafter\ifx\csname urlprefix\endcsname\relax\def\urlprefix{URL }\fi
\expandafter\ifx\csname doiprefix\endcsname\relax\def\doiprefix{DOI }\fi
\providecommand{\bibinfo}[2]{#2}
\providecommand{\eprint}[2][]{\url{#2}}

\bibitem{MCTS1}
\bibinfo{author}{Kunz-Schugart, L.~A.}, \bibinfo{author}{Kreutz, M.} \&
  \bibinfo{author}{Kneuchel, R.}
\newblock \bibinfo{journal}{\bibinfo{title}{Multicellular spheroids: a
  three-dimensional in vitro culture system to study tumour biology}}.
\newblock {\emph{\JournalTitle{International Journal of Experimental
  Pathology}}} \textbf{\bibinfo{volume}{79}}, \bibinfo{pages}{1 -- 23}
  (\bibinfo{year}{1998}).

\bibitem{casciari1992variations}
\bibinfo{author}{Casciari, J.~J.}, \bibinfo{author}{Sotirchos, S.~V.} \&
  \bibinfo{author}{Sutherland, R.~M.}
\newblock \bibinfo{journal}{\bibinfo{title}{Variations in tumor cell growth
  rates and metabolism with oxygen concentration, glucose concentration, and
  extracellular ph}}.
\newblock {\emph{\JournalTitle{Journal of Cellular Physiology}}}
  \textbf{\bibinfo{volume}{151}}, \bibinfo{pages}{386--394}
  (\bibinfo{year}{1992}).

\bibitem{mueller2000tumor}
\bibinfo{author}{Mueller-Klieser, W.}
\newblock \bibinfo{journal}{\bibinfo{title}{Tumor biology and experimental
  therapeutics}}.
\newblock {\emph{\JournalTitle{Critical Reviews in Oncology/Hematology}}}
  \textbf{\bibinfo{volume}{36}}, \bibinfo{pages}{123--139}
  (\bibinfo{year}{2000}).

\bibitem{targ1}
\bibinfo{author}{Wilson, W.~R.} \& \bibinfo{author}{Hay, M.~P.}
\newblock \bibinfo{journal}{\bibinfo{title}{Targeting hypoxia in cancer
  therapy}}.
\newblock {\emph{\JournalTitle{Nature Reviews Cancer}}}
  \textbf{\bibinfo{volume}{11}}, \bibinfo{pages}{393 -- 410}
  (\bibinfo{year}{2011}).

\bibitem{Mao2}
\bibinfo{author}{Mao, X.} \emph{et~al.}
\newblock \bibinfo{journal}{\bibinfo{title}{An agent-based model for
  drug-radiation interactions in the tumour microenvironment: Hypoxia-activated
  prodrug sn30000 in multicellular tumour spheroids}}.
\newblock {\emph{\JournalTitle{PLOS Computational Biology}}}
  \textbf{\bibinfo{volume}{14}}, \bibinfo{pages}{1--30} (\bibinfo{year}{2018}).
\newblock \doiprefix 10.1371/journal.pcbi.1006469.

\bibitem{Diaspro:02}
\bibinfo{author}{Diaspro, A.}, \bibinfo{author}{Federici, F.} \&
  \bibinfo{author}{Robello, M.}
\newblock \bibinfo{journal}{\bibinfo{title}{Influence of refractive-index
  mismatch in high-resolution three-dimensional confocal microscopy}}.
\newblock {\emph{\JournalTitle{Applied Optics}}} \textbf{\bibinfo{volume}{41}},
  \bibinfo{pages}{685--690} (\bibinfo{year}{2002}).

\bibitem{hong2018cellular}
\bibinfo{author}{Hong, C.~R.} \emph{et~al.}
\newblock \bibinfo{journal}{\bibinfo{title}{Cellular pharmacology of
  evofosfamide (th-302): A critical re-evaluation of its bystander effects}}.
\newblock {\emph{\JournalTitle{Biochemical pharmacology}}}
  \textbf{\bibinfo{volume}{156}}, \bibinfo{pages}{265--280}
  (\bibinfo{year}{2018}).

\bibitem{welch2011optical}
\bibinfo{author}{Welch, A.~J.} \& \bibinfo{author}{Van~Gemert, M.~J.}
\newblock \emph{\bibinfo{title}{Optical-thermal response of laser-irradiated
  tissue}}, vol.~\bibinfo{volume}{2} (\bibinfo{publisher}{Springer},
  \bibinfo{year}{2011}).

\bibitem{sharma2007imaging}
\bibinfo{author}{Sharma, M.}, \bibinfo{author}{Verma, Y.},
  \bibinfo{author}{Rao, K.}, \bibinfo{author}{Nair, R.} \&
  \bibinfo{author}{Gupta, P.}
\newblock \bibinfo{journal}{\bibinfo{title}{Imaging growth dynamics of tumour
  spheroids using optical coherence tomography}}.
\newblock {\emph{\JournalTitle{Biotechnology letters}}}
  \textbf{\bibinfo{volume}{29}}, \bibinfo{pages}{273--278}
  (\bibinfo{year}{2007}).

\bibitem{huang2017optical}
\bibinfo{author}{Huang, Y.} \emph{et~al.}
\newblock \bibinfo{journal}{\bibinfo{title}{Optical coherence tomography
  detects necrotic regions and volumetrically quantifies multicellular tumor
  spheroids}}.
\newblock {\emph{\JournalTitle{Cancer research}}} \bibinfo{pages}{canres--0821}
  (\bibinfo{year}{2017}).

\bibitem{LiuRefractive}
\bibinfo{author}{Liu, P.~Y.} \emph{et~al.}
\newblock \bibinfo{journal}{\bibinfo{title}{Cell refractive index for cell
  biology and disease diagnosis: past{,} present and future}}.
\newblock {\emph{\JournalTitle{Lab on a Chip}}} \textbf{\bibinfo{volume}{16}},
  \bibinfo{pages}{634--644} (\bibinfo{year}{2016}).

\bibitem{Tearney:95}
\bibinfo{author}{Tearney, G.~J.} \emph{et~al.}
\newblock \bibinfo{journal}{\bibinfo{title}{Determination of the refractive
  index of highly scattering human tissue by optical coherence tomography}}.
\newblock {\emph{\JournalTitle{Optics Letters}}} \textbf{\bibinfo{volume}{20}},
  \bibinfo{pages}{2258--2260} (\bibinfo{year}{1995}).

\bibitem{Uhlhorn20082732}
\bibinfo{author}{Uhlhorn, S.~R.}, \bibinfo{author}{Borja, D.},
  \bibinfo{author}{Manns, F.} \& \bibinfo{author}{Parel, J.-M.}
\newblock \bibinfo{journal}{\bibinfo{title}{Refractive index measurement of the
  isolated crystalline lens using optical coherence tomography}}.
\newblock {\emph{\JournalTitle{Vision Research}}}
  \textbf{\bibinfo{volume}{48}}, \bibinfo{pages}{2732--2738}
  (\bibinfo{year}{2008}).

\bibitem{Mengteeth}
\bibinfo{author}{Meng, Z.} \emph{et~al.}
\newblock \bibinfo{journal}{\bibinfo{title}{Measurement of the refractive index
  of human teeth by optical coherence tomography}}.
\newblock {\emph{\JournalTitle{Journal of Biomedical Optics}}}
  \textbf{\bibinfo{volume}{14}}, \bibinfo{pages}{034010--034010--4}
  (\bibinfo{year}{2009}).

\bibitem{Chakraborty201462}
\bibinfo{author}{Chakraborty, R.}, \bibinfo{author}{Lacy, K.~D.},
  \bibinfo{author}{Tan, C.~C.}, \bibinfo{author}{Park, H.~n.} \&
  \bibinfo{author}{Pardue, M.~T.}
\newblock \bibinfo{journal}{\bibinfo{title}{Refractive index measurement of the
  mouse crystalline lens using optical coherence tomography}}.
\newblock {\emph{\JournalTitle{Experimental Eye Research}}}
  \textbf{\bibinfo{volume}{125}}, \bibinfo{pages}{62 -- 70}
  (\bibinfo{year}{2014}).

\bibitem{lippok2012dispersion}
\bibinfo{author}{Lippok, N.}, \bibinfo{author}{Coen, S.},
  \bibinfo{author}{Nielsen, P.} \& \bibinfo{author}{Vanholsbeeck, F.}
\newblock \bibinfo{journal}{\bibinfo{title}{Dispersion compensation in fourier
  domain optical coherence tomography using the fractional fourier transform}}.
\newblock {\emph{\JournalTitle{Optics express}}} \textbf{\bibinfo{volume}{20}},
  \bibinfo{pages}{23398--23413} (\bibinfo{year}{2012}).

\bibitem{MaxGitHub}
\bibinfo{author}{Maske, M.}
\newblock \bibinfo{title}{{S}pheroid {RGB}}.
\newblock \bibinfo{howpublished}{\url{https://github.com/IamMM/Spheroid_RGB}}.

\bibitem{zhang2016optimization}
\bibinfo{author}{Zhang, W.} \emph{et~al.}
\newblock \bibinfo{journal}{\bibinfo{title}{Optimization of the formation of
  embedded multicellular spheroids of mcf-7 cells: How to reliably produce a
  biomimetic 3d model}}.
\newblock {\emph{\JournalTitle{Analytical biochemistry}}}
  \textbf{\bibinfo{volume}{515}}, \bibinfo{pages}{47--54}
  (\bibinfo{year}{2016}).

\bibitem{martelli2004effect}
\bibinfo{author}{Martelli, F.}, \bibinfo{author}{Del~Bianco, S.} \&
  \bibinfo{author}{Zaccanti, G.}
\newblock \bibinfo{journal}{\bibinfo{title}{Effect of the refractive index
  mismatch on light propagation through diffusive layered media}}.
\newblock {\emph{\JournalTitle{Physical Review E}}}
  \textbf{\bibinfo{volume}{70}}, \bibinfo{pages}{011907}
  (\bibinfo{year}{2004}).

\bibitem{BistaRI2011}
\bibinfo{author}{Bista, R.~K.} \emph{et~al.}
\newblock \bibinfo{journal}{\bibinfo{title}{Quantification of nanoscale nuclear
  refractive index changes during the cell cycle}}.
\newblock {\emph{\JournalTitle{Journal of Biomedical Optics}}}
  \textbf{\bibinfo{volume}{16}}, \bibinfo{pages}{070503--070503--3}
  (\bibinfo{year}{2011}).

\bibitem{liang2007determining}
\bibinfo{author}{Liang, X.~J.}, \bibinfo{author}{Liu, A.~Q.},
  \bibinfo{author}{Lim, C.~S.}, \bibinfo{author}{Ayi, T.~C.} \&
  \bibinfo{author}{Yap, P.~H.}
\newblock \bibinfo{journal}{\bibinfo{title}{Determining refractive index of
  single living cell using an integrated microchip}}.
\newblock {\emph{\JournalTitle{Sensors and Actuators A: Physical}}}
  \textbf{\bibinfo{volume}{133}}, \bibinfo{pages}{349--354}
  (\bibinfo{year}{2007}).

\bibitem{Choi10}
\bibinfo{author}{Choi, W.~J.} \emph{et~al.}
\newblock \bibinfo{journal}{\bibinfo{title}{Full-field optical coherence
  microscopy for identifying live cancer cells by quantitative measurement of
  refractive index distribution}}.
\newblock {\emph{\JournalTitle{Optics Express}}} \textbf{\bibinfo{volume}{18}},
  \bibinfo{pages}{23285--23295} (\bibinfo{year}{2010}).

\bibitem{Tannock878}
\bibinfo{author}{Tannock, I.~F.}, \bibinfo{author}{Lee, C.~M.},
  \bibinfo{author}{Tunggal, J.~K.}, \bibinfo{author}{Cowan, D. S.~M.} \&
  \bibinfo{author}{Egorin, M.~J.}
\newblock \bibinfo{journal}{\bibinfo{title}{Limited penetration of anticancer
  drugs through tumor tissue}}.
\newblock {\emph{\JournalTitle{Clinical Cancer Research}}}
  \textbf{\bibinfo{volume}{8}}, \bibinfo{pages}{878--884}
  (\bibinfo{year}{2002}).

\bibitem{Mao1}
\bibinfo{author}{Mao, X.}, \bibinfo{author}{McManaway, S.},
  \bibinfo{author}{Wilson, W.~R.} \& \bibinfo{author}{Hicks, K.~O.}
\newblock \bibinfo{journal}{\bibinfo{title}{Abstract 2440: Potentiation of the
  action of chemotherapeutic drugs by the hypoxia-activated prodrug sn30000 in
  multicellular tumor spheroids}}.
\newblock {\emph{\JournalTitle{Cancer Research}}}
  \textbf{\bibinfo{volume}{78}}, \bibinfo{pages}{2440--2440}
  (\bibinfo{year}{2018}).
\newblock \doiprefix 10.1158/1538-7445.AM2018-2440.

\bibitem{vermes1994apoptosis}
\bibinfo{author}{Vermes, I.} \& \bibinfo{author}{Haanen, C.}
\newblock \bibinfo{journal}{\bibinfo{title}{Apoptosis and programmed cell death
  in health and disease}}.
\newblock {\emph{\JournalTitle{Advances in Clinical Chemistry}}}
  (\bibinfo{year}{1994}).

\bibitem{kerr1972apoptosis}
\bibinfo{author}{Kerr, J.~F.}, \bibinfo{author}{Wyllie, A.~H.} \&
  \bibinfo{author}{Currie, A.~R.}
\newblock \bibinfo{journal}{\bibinfo{title}{Apoptosis: a basic biological
  phenomenon with wide-ranging implications in tissue kinetics}}.
\newblock {\emph{\JournalTitle{British journal of cancer}}}
  \textbf{\bibinfo{volume}{26}}, \bibinfo{pages}{239} (\bibinfo{year}{1972}).

\bibitem{van2010apoptosis}
\bibinfo{author}{van~der Meer, F.~J.} \emph{et~al.}
\newblock \bibinfo{journal}{\bibinfo{title}{Apoptosis-and necrosis-induced
  changes in light attenuation measured by optical coherence tomography}}.
\newblock {\emph{\JournalTitle{Lasers in medical science}}}
  \textbf{\bibinfo{volume}{25}}, \bibinfo{pages}{259--267}
  (\bibinfo{year}{2010}).

\end{thebibliography}

\section*{Acknowledgements}
The authors would like to thank the Marsden Fund (UoA1509, 13-UOA-187) from the Royal Society of New Zealand, the Health Research Council of New Zealand (14/538), and the Dodd-Walls Centre for Photonics and Quantum Technologies for their support. \\
The authors would like to also thank Dr Bastian Braeuer for his aid with the experiments and to Maximillian Maske for developing the SpheroidRGB plugin for ImageJ.


\section*{Author contributions statement}

N.H. conducted the OCT measurements. P.P. prepared the spheroids and conducted the confocal microscopy measurements. F.V., J.R., K.H. conceived the experiments and supervised the work.  All authors contributed to the results analysis, wrote, and reviewed the manuscript. 

\section*{Additional information}

\textbf{Competing interests}
The authors declare no competing interests.

\textbf{Data availability}
The datasets generated during and analysed during the current study are available from the corresponding author on reasonable request.

\end{document}